\documentclass[11pt,a4paper]{article}
\usepackage{heppub}
\pdfoutput=1
\usepackage[x11names]{xcolor}
\usepackage{subcaption}
\usepackage[T1]{fontenc}
\usepackage[utf8]{inputenc}
\usepackage{braket}
\usepackage{slashed}
\usepackage{placeins}
\usepackage{multirow}
\usepackage[normalem]{ulem}
\usepackage{tikz}
\usetikzlibrary{calc}
\usepackage{colortbl}
\usepackage{makecell}

\newcommand{\MSbar}{{\overline{\mathrm{MS}}}}
\newcommand{\df}{\mathrm{d}}
\newcommand{\img}{\mathrm{i}} 
\newcommand{\eps}{\epsilon}
\newcommand{\cA}{\mathcal{A}}

\newcommand{\nn}{\nonumber}

\newcommand{\UV}{\mathrm{UV}}

\newcommand{\Tr}{\mathrm{Tr}}

\newcommand{\bt}{\vec b_T}

\newcommand{\pt}{\vec p_T}

\newcommand{\uv}{\mathrm{uv}}

\newcommand{\Corr}{\Omega}

\newcommand{\na}{n_a}
\newcommand{\nb}{n_b}
\newcommand{\nA}{n_A}
\newcommand{\nB}{n_B}

\newcommand{\LR}{\mathrm{LR}}

\newcommand{\staple}{\sqsupset}
\newcommand\softstaple[1][1]
{
	\begin{tikzpicture}[scale=#1]
		\coordinate (dx) at (0.15, 0);
		\coordinate (dy) at (0   , 0.1);
		\coordinate (bT) at (0.  , 0.05);
		\draw [line width=0.15mm] (0,0) -- ($(dx) + (dy)$) -- ($2*(dy)$) -- ($2*(dy)+(bT)$) -- ($(dx) + (dy) + (bT)$) -- ($(bT)$) -- (0,0);
		\draw [line width=0.12mm] (0,0) -- (0, -0.0055);
	\end{tikzpicture}
}

\begin{document}
	

\title{One-loop matching for gluon lattice TMDs}
\author[a]{Stella T.~Schindler,}
\author[a]{Iain W.~Stewart,}
\author[b]{and Yong Zhao}

\affiliation[a]{Center for Theoretical Physics,\,Massachusetts Institute of Technology,\,Cambridge,\,MA\,02139,\,USA}
\affiliation[b]{Physics Division, Argonne National Laboratory, Lemont, IL 60439, USA}

\emailAdd{stellas@mit.edu}
\emailAdd{iains@mit.edu}
\emailAdd{yong.zhao@anl.gov}

\abstract{
Transverse-momentum-dependent parton distributions (TMDs) can be calculated from first principles by computing a related set of Euclidean lattice observables and connecting them via a factorization formula.
This work focuses on the leading-power factorization formula connecting the lattice quasi-TMD and continuum Collins TMD for gluons. We calculate the one-loop gluon matching coefficient, which is known to be independent of spin and exhibits no mixing with quarks. 
We demonstrate that this coefficient satisfies Casimir scaling with respect to the quark matching coefficient at one-loop order. Our result facilitates reliable lattice QCD calculations of gluon TMDs.
}

\preprint{\vbox{
		\hbox{MIT--CTP 5427}
}}

\arxivnumber{}

\maketitle

\allowdisplaybreaks

\section{Introduction}

Protons are a fundamental building block of almost all the visible matter in the universe, yet over a century after their discovery, their internal dynamics still confound us. 
The behavior of quarks and gluons inside the proton are described by universal functions appearing in cross sections, such as parton distribution functions (PDFs), tranverse-momentum-dependent PDFs (TMDs), and generalized parton distributions (GPDs).
Various collider facilities, such as Jefferson Lab, CERN, DESY, and Fermilab, have provided initial data on these distributions, and many have dedicated plans to increase their precision. 
These efforts will be further bolstered by the planned Electron-Ion Collider (EIC) \cite{Accardi:2012qut, AbdulKhalek:2021gbh} at Brookhaven National Lab.
The imminence of high-precision data about parton distributions underscores the need for developing a corresponding first-principles understanding. 

Parton distributions can be rigorously defined in quantum field theory.
Unfortunately, they always contain important nonperturbative components that cannot be computed directly using lattice QCD, as they are typically described by matrix elements involving light-like Wilson line paths.
Light-like observables depend on a real-valued time variable, which induces a lattice sign problem: a numerical difficulty suspected to be NP-hard in general. 
Absent direct analytical or numerical approaches, physicists have traditionally constructed
models for such observables, which are then compared to experimental data using global fits. 
Alternatively, one can carry out an \emph{indirect} first-principles calculation: 
one constructs a numerically-tractable lattice observable that shares the same IR physics as the target distribution, but may differ in the ultraviolet (UV). 
This lattice observable must then be perturbatively matched back onto the desired physical continuum distribution via a factorization formula.

Here, we focus on the physical-to-lattice factorization program for TMDs, an important class of observables describing the three-dimensional momentum distribution of quarks and gluons inside a proton. 
TMDs appear in many cross sections, including semi-inclusive deep-inelastic scattering (SIDIS), the Drell-Yan process, weak boson production, and Higgs production. 
Nonperturbative effects for TMDs dominate at small transverse momenta $k_T\sim \Lambda_{\rm QCD}$. 
For $k_T\gg \Lambda_{\rm QCD}$, the dependence of the TMD on $k_T$ can be computed perturbatively~\cite{Collins:1981va,Aybat:2011zv, Catani:2012qa, Collins:2011zzd, Gehrmann:2014yya,Echevarria:2016scs, Luebbert:2016itl, Luo:2019hmp, Luo:2019szz, Luo:2020epw, Ebert:2020yqt}; but there are still important nonperturbative contributions~\cite{Bacchetta:2017gcc,Scimemi:2018xaf,Lustermans:2019plv,Ebert:2022cku}.
On the modeling front, high precision global fits with next-to-next-to-next-to-leading logarithmic accuracy on perturbative ingredients have been performed for the existing experimental data \cite{Scimemi:2019cmh, Bacchetta:2019sam}.
On the lattice front, the calculation of TMD-like correlators was pioneered and developed over the course of a decade using the Musch-H{\"a}gler-Engelhardt-Negele-Sch{\"a}fer (MHENS) scheme \cite{Hagler:2009mb, Musch:2010ka, Musch:2011er, Engelhardt:2015xja, Yoon:2016dyh, Yoon:2017qzo}. 
Later on, quasi-TMDs were developed for the lattice~\cite{Ji:2014hxa,Ji:2018hvs,Ebert:2018gzl,Ebert:2019okf,Ebert:2019tvc,Ji:2019sxk,Ji:2019ewn,Ebert:2020gxr,Ji:2020jeb,Ji:2021znw} using the framework of large momentum effective theory (LaMET)~\cite{Ji:2013dva,Ji:2014gla,Ji:2020ect}. 
The leading-power factorization formula connecting quasi-TMDs to the continuum has been proposed in Refs.~\cite{Ebert:2018gzl,Ebert:2019okf,Ji:2019sxk,Ji:2019ewn}, and recently derived to all orders in $\alpha_s$ in~\cite{Ebert:2022fmh}.
First lattice results have been presented for various components of quasi-TMDs: the  beam function and Collins-Soper kernel \cite{Shanahan:2019zcq, Shanahan:2020zxr, Schlemmer:2021aij, LatticeParton:2020uhz, Li:2021wvl, Shanahan:2021tst}, and soft function \cite{LatticeParton:2020uhz, Li:2021wvl}.

At leading power, the factorization formula connecting the continuum limit of quasi-TMDs (lattice) and the physical Collins TMD reads~\cite{Ebert:2022fmh}: 
\begin{align} \label{eq:factorization_statement}
	\tilde f_{i/h}^{[s]}(x, \bt, \mu, \tilde\zeta, x\tilde P^z)
	&= C_{\kappa_i}(x \tilde P^z, \mu)
	\exp\biggl[ \frac12 \gamma_\zeta^{\kappa_i}(\mu, b_T) \ln\frac{\tilde\zeta }{\zeta}\biggr]
	f_{i/h}^{[s]}(x, \bt, \mu, \zeta)
\,,\end{align}
where $f$ is a continuum TMD, $\tilde{f}$ is the continuum limit of a lattice TMD, $C_{\kappa_i}$ is a perturbative matching coefficient, and the exponential term evolves the so-called Collins-Soper (CS) scale $\zeta\to \tilde{\zeta}$ using the CS kernel $\gamma_\zeta^{\kappa_i}$.
The other parameters include $x$, which is the fraction of the hadron $h$'s longitudinal momentum $P$ that the parton of type $i$ carries; the Fourier-conjugate to the parton's transverse momentum $b_T$; and a renormalization scale $\mu$. 
Here $i$ refers to either a gluon ($i=g$) or specific quark flavor ($i=u,d,s,\ldots$), where $\kappa_i=q$ is universal for all quarks, but differs from $\kappa_g=g$ for gluons.
No quark-gluon or flavor mixing occurs in the factorization relation, which simplifies calculation of gluon TMDs from the lattice \cite{Ebert:2022fmh}.
The definition of each of the TMDs in \eq{factorization_statement} will be discussed in detail in \sec{defs}. 
 
The quasi-to-Collins matching coefficient $C_q$ is known for quarks at one loop~\cite{Ji:2014hxa,Ebert:2018gzl,Ebert:2019okf}, for logarithmic terms at two loops~\cite{Ji:2019ewn}, and for all next-to-leading-logarithmic (NLL) terms~\cite{Ebert:2022fmh}.
A key missing ingredient is the matching coefficient for gluons. Gluons are known to be responsible for a large part of internal proton dynamics and would be useful to understand from first principles.
In this work, we calculate the one-loop gluon matching coefficient $C_g$.

\subsection{Definitions of lattice and physical continuum TMDs}
\label{sec:defs}

\begin{figure*}
	\centering
	\raisebox{4ex}{\includegraphics[width=0.5\textwidth]{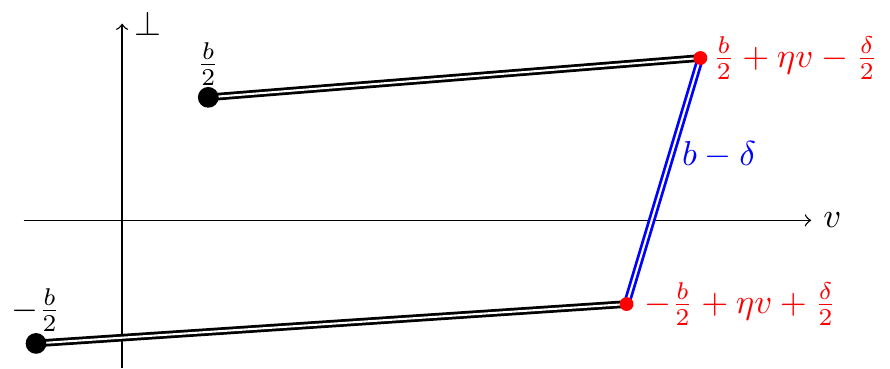}}
	\hspace{1cm}
	\includegraphics[width=0.4\textwidth]{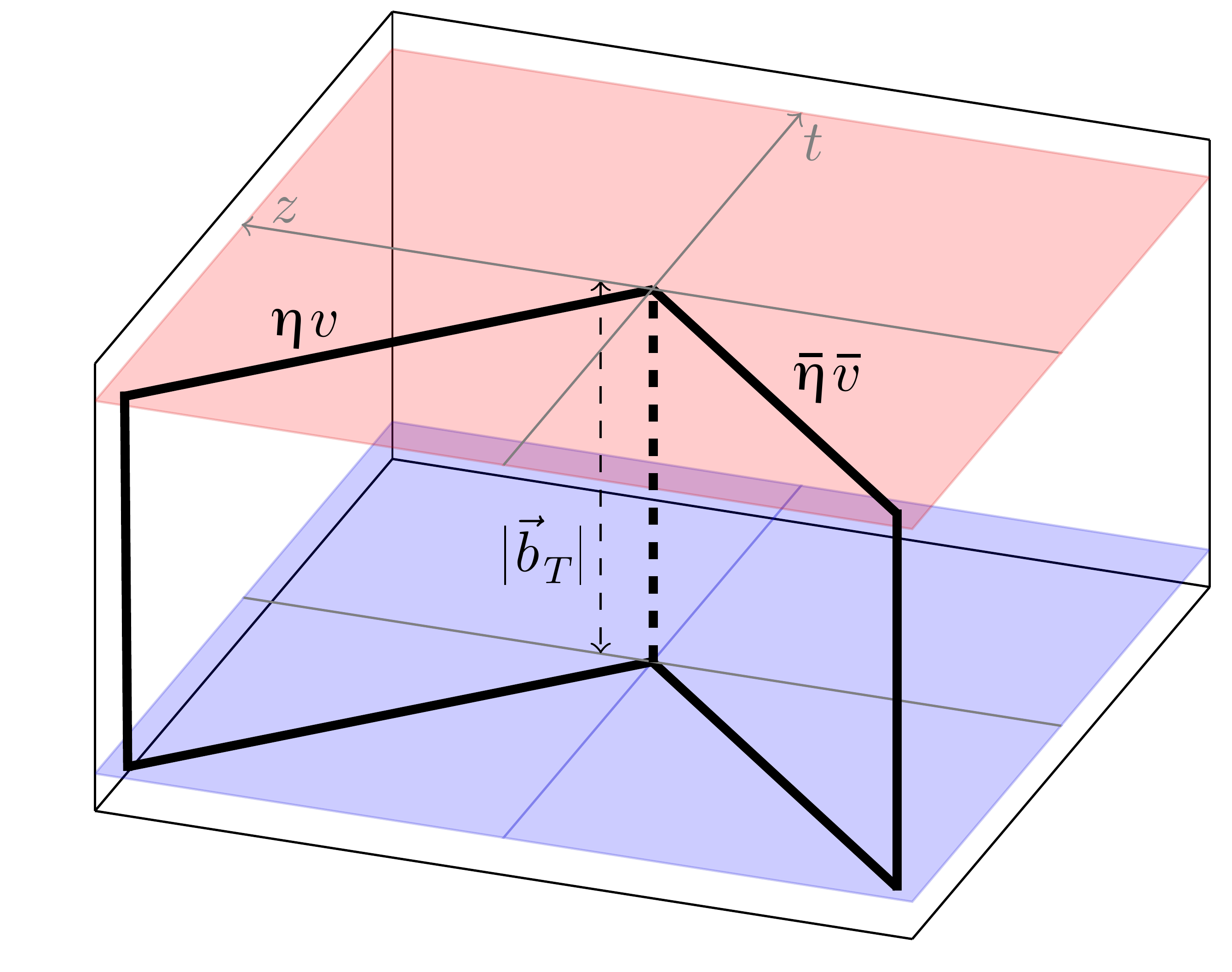}
	\caption{(a) Staple-shaped Wilson line defined in \eq{W_staple}.
		Edges may extend along the conjugate direction $P$, which is not shown.
		(b) Wilson line structure of the soft function, \eq{soft_generic} for $\eta,\bar\eta<0$. Figure (a) is adapted from \refcite{Ebert:2022fmh} and (b) is adapted from \refcite{Li:2016axz}.}
	\label{fig:generic_staples}
\end{figure*}

A TMD $f_{i/h}$ for gluon or quark flavor $i$ generally is a product of three components:
a beam function (hadronic matrix element), a soft function (vacuum matrix element), and their renormalization.
The beam and soft matrix elements are defined in terms of Wilson lines:
\begin{align}
	W^R[\gamma]
	= P \exp\biggl[ \img g \int_{\gamma} \df x^\mu \cA_\mu^a(x)\, T_R^a\biggr]
	\,,\end{align}
where $\gamma$ is the Wilson line's path and $R$ is its color representation.
Here we focus on gluons, taking $R=g$ for the adjoint representation.
Beam functions involve staple-shaped Wilson line paths as shown in \fig{generic_staples},
\begin{align} \label{eq:W_staple}
	W_\staple^g(b, \eta v, \delta)
	= W^g\left[ \frac{b}{2} ~\to~ \frac{b}{2} + \eta v - \frac{\delta}{2}
	~\to~ -\frac{b}{2} + \eta v + \frac{\delta}{2}
	~\to -\frac{b}{2} \right]
	\,.\end{align}
A generic notation for gluon beam functions is given by
\begin{align} \label{eq:beam_generic}
	\Corr_{g/h}^{\mu\nu\rho\sigma}(b, P, \eps, \eta v, \delta) &
	= \Bigl\langle h(P) \Big| G^{\mu\nu}\Bigl(\frac{b}{2}\Bigr)
	W^g_\staple(b, \eta v, \delta)
	G^{\rho\sigma}\Bigl(-\frac{b}{2}\Bigr) \Big| h(P) \Bigr\rangle
	\,.\end{align}
Here, $G^{\mu\nu}(b)$ is the gluon field strength tensor.
Gluon fields are spatially separated by $b$, which is Fourier-conjugate to the struck parton's momentum, $h$ denotes a parent hadron with momentum $P$,
$\eps$ is the UV regulator, and $\eta v$ and $\delta$ characterize the Wilson line path.
See \refscite{Mulders:2000sh,Echevarria:2015uaa} for the decomposition of gluon TMDs into independent spin structures.

Note that the correlator $\Omega$ characterizes the TMDs in $b$-space.
When we define a full TMD in terms of beam and soft functions, we generally make a Fourier transform  $P\cdot b \to x$ and write
\begin{equation}
	\Omega(b,  ... ) \to B(x, \bt, ... ),
\end{equation}
where $B$ indicates the correlator in these new coordinates.
	
The gluon soft function is defined as
\begin{align} \label{eq:soft_generic}
	S^g\left[b, \eps, \eta v, \bar\eta \bar v\right] &
	= \frac{1}{N_c^2-1} \Big\langle 0 \Big| \Tr \Bigl[ S^g_{\softstaple}(b, \eta v, \bar\eta \bar v) \Bigr] \Big| 0 \Bigr\rangle
	\,,\end{align}
where the trace is over color, and the soft Wilson line is 
\begin{align} \label{eq:S_staple}
	S^g_{\softstaple}(b, \eta v, \bar\eta \bar v) &
	= W^g\bigg[\frac{b}{2} \to \frac{b}{2}+ \bar\eta \bar v \to -\frac{b}{2} + \bar\eta \bar v \to -\frac{b}{2}
	\to -\frac{b}{2} + \eta v ~\to~ \frac{b}{2} + \eta v \to \frac{b}{2} \bigg]
	\,,\end{align}
as shown in \fig{generic_staples}.
The Wilson line $S_{\softstaple}$ consists of two beam function staples glued together at the points $\pm b/2$; the long sides of the staples run along the $\bar \eta \bar v$ and $\eta v$ directions.   
Note that the figure shows the case where $b=\vec b_T$, and the points are translated to $0$ and $-\vec b_T$.

We can now define all off-lightcone gluon TMD schemes in terms of the exact same $B$ and $S$: 
\begin{equation} \label{eq:TMDdef}
	f_{g/h}^{[s]}(x,\vec{b}_T, \mu, P, v, \bar{v}, \delta, \ldots) = \lim_{\substack{\rm  lightcone\\ \rm \&\: UV\:limits}} Z^g_{\UV}(\eps,\mu,v, \bar{v}, \delta, \ldots)  \lim_{\eta\to\infty} \frac{B_{g/h}^{\rho\sigma}(x,\bt, P, \eps, \eta v, \delta) }{\sqrt{S^g\left[b, \eps, \eta v, \bar\eta \bar v\right] }},
\end{equation}
where the ellipses and $Z^g_{\UV}$ account for a scheme-dependent renormalization.~\footnote{Note that for the MHENS scheme, renormalization should be performed before the Fourier transform as $Z^g_{\UV}$ depends on the longitudinal component of $b^\mu$.}
Each scheme is characterized by a specific choice of the parameters $b^\mu$,  $\eta v^\mu$,  $P^\mu$, and $\delta^\mu$ appearing in the beam and soft functions, as well as the precise nature and order of the lightcone and UV limits in \eq{TMDdef}.
The longitudinal staple length $\eta$ distinguishes physical TMDs ($\eta = \pm \infty$) from lattice TMDs (finite $\eta$).

The remainder of \sec{defs} specifies the order of limits and beam/soft function arguments that define three major TMD schemes: the continuum Collins TMD, the lattice quasi-TMD, and the continuum Large Rapidity (LR) TMD. 
We summarize this information in table \ref{tbl:TMDs}, which also includes values taken on by the four-vectors $b^\mu,\,v^\mu,\,\delta^\mu,\,P^\mu$.  

\paragraph{Collins scheme~\cite{Collins:2011zzd}.}
The continuum Collins scheme uses spacelike Wilson lines with directions
\begin{alignat}{2} \label{eq:nb}
	\nA^\mu(y_A) &\equiv \na^\mu - e^{-2y_A} \nb^\mu &&= (1, -e^{-2 y_A}, 0_\perp)
	\,,\nn\\
	\nB^\mu(y_B) &\equiv \nb^\mu - e^{2y_B} \na^\mu &&= (-e^{2 y_B}, 1, 0_\perp)
	\,,\end{alignat}
parametrized by the rapidities $y_A$ and $y_B$. {(See \app{notation} for our lightcone coordinate conventions.)}
The Collins gluon TMD for a hadron $h$ moving along $n_a$ with rapidity $y_P$ is
\begin{align} \label{eq:tmdpdf_Collins}
	& f_{g/h}(x, \bt, \mu, \zeta)
	= \lim_{\eps\to0} Z_\uv^g(\eps, \mu, \zeta)
	\lim_{\substack{y_B \to -\infty}}
	\frac{B_{g/h}^{C\rho\sigma}(x, \bt, \eps, y_P - y_B)}{\sqrt{S^g \left[b_\perp,\eps,-\infty n_A(2y_n),-\infty n_B(2y_B)\right]}}
	\,,\end{align}
where $|b_\perp|=b_T$ and the beam function is defined as
\begin{align}
 B_{g/h}^{C\rho\sigma}(x, \bt, \eps, y_P - y_B) &
= \int\!\frac{\df b^-}{2\pi} \frac{e^{-\img b^- (x P^+)}}{x P^+}
\Corr_{g/h}^{+\rho+\sigma}\bigl[b, P, \eps, -\infty \nB(y_B), b^- n_b  \bigr]\,,
\end{align}
with $\Corr_{g/h}^{+\rho+\sigma} = (n_b)_{\mu}\ \Corr_{g/h}^{\mu\rho\nu\sigma}\	(n_b)_\nu$.
The CS scale $\zeta$ is defined as
\begin{align} \label{eq:zeta}
	\zeta = 2 (x P^+ e^{-y_n})^2 = x^2 m_h^2 e^{2(y_P - y_n)}\,.
\end{align}

\paragraph{Quasi-TMDs~\cite{Ji:2014hxa,Ji:2018hvs,Ebert:2018gzl,Ebert:2019okf,Ebert:2019tvc,Ji:2019sxk,Ji:2019ewn,Ebert:2020gxr,Ji:2020jeb,Ji:2021znw,Ebert:2022fmh}.} 
Collins TMDs cannot be directly discretized and computed on the lattice, as their Wilson lines' explicit dependence on time induces a sign problem. 
 	Instead, one defines numerically-tractable quasi-TMD beam functions using finite-length Wilson lines lying on equal-time paths. The lattice gluon quasi-TMD takes the form
\begin{align}\label{eq:qTMDdef}
 \tilde f_{i/h}^{[\tilde \Gamma]} (x, \bt, \mu, \tilde \zeta, x\tilde P^z)
& = \lim_{\substack{\tilde \eta\to\infty \\ a\to0}} \,
   \frac{Z'^g_{\rm uv}(\mu,\tilde \mu)
    Z^g_{\rm uv}(a, \tilde\mu,y_n\!-\!y_B) \tilde B_{g/h}^{\alpha\rho\beta\sigma}(x, \bt, a, \tilde\eta, x\tilde P^z)}
    {\sqrt{S^g\left[b_\perp,a,-\tilde\eta\,\cfrac{ n_A(2y_n)}{|n_A(2y_n)|},-\tilde\eta\,\cfrac{ n_B(2y_B)}{|n_B(2y_B)|} \right]}}
\,,\end{align}
where the bare quasi-beam function reads as
\begin{align} \label{eq:qbeam}
 \tilde B_{g/h}^{\alpha\rho\beta\sigma}(x, \bt, a, \tilde\eta, x \tilde P^z) &
 = N^{\alpha\rho\beta\sigma} \int \frac{\df \tilde b^z}{2\pi} \, \frac{e^{\img \tilde b^z (x \tilde P^z)}}{x \tilde P^z}
   \,\Corr_{g/h}^{\alpha\rho\beta\sigma}\left[\tilde b, \tilde P, a, \tilde\eta \hat z, \tilde b^z \hat z\right]
\,.\end{align}
Here, $\alpha$, $\beta$, $\rho$ and $\sigma$ are generic Lorentz indices,
and $N^{\alpha\rho\beta\sigma}$ is a trivial normalization factor. The factor $Z^g_{\rm uv}(a, \tilde\mu,y_n-y_B)$ renormalizes UV divergences from lattice regularization, and $Z'^g_{\rm uv}(\mu,\tilde \mu)$ matches the result to the $\MSbar$ scheme.
The second equality in \eq{qTMDdef} holds for large $\tilde P^z$, and the variable $\tilde\zeta$ is
\begin{align}
\tilde \zeta = \big(x m_h e^{ y_{\tilde P}+y_B  - y_n}\big)^2 \approx (2 x \tilde P^z e^{y_B-y_n})^2
\,.\end{align}

{
	\renewcommand{\arraystretch}{2.6}
	\begin{table*}
		\centering
		\setlength{\tabcolsep}{1em}
		\fontsize{9}{9}\selectfont
		\begin{tabular}{|c|c|c|}
			\cellcolor{Snow3} & \cellcolor{Snow3} Collins TMD (continuum)& \cellcolor{Snow3} Quasi-TMD (lattice)
			\\ \Xhline{3\arrayrulewidth}
			\cellcolor{Snow3} TMD & $\lim\limits_{\eps\to 0} Z^{\kappa_i}_{\text{UV}} \lim\limits_{y_B\to -\infty}\cfrac{B_{i/h}} {\sqrt{S^{\kappa_i}}}$ &  $\lim\limits_{a\to 0}Z_{\text{UV}}^{\kappa_i}\cfrac{\tilde{B}_{i/h}}{\sqrt{ S^{\kappa_i}}}$
			\\ \hline
			\cellcolor{Snow3} Beam function 
           & $\Corr_{i/h}\left[b,P,\eps,-\infty n_B(y_B),b^-n_b\right] \stackrel{\rm FT}{\longrightarrow} B_{i/h}$ 
           & $\Corr_{i/h}(\tilde b,\tilde P,a,\tilde\eta \hat{z},\tilde b^z\hat{z})\stackrel{\rm FT}{\longrightarrow} \tilde B_{i/h}$
			\\ \hline
			\cellcolor{Snow3} Soft function & $S^{\kappa_i} \left[b_\perp,\eps,-\infty n_A(2y_n),-\infty n_B(2y_B)\right]$ & $S^{\kappa_i}\left[b_\perp,a,-\tilde\eta\cfrac{n_A(2y_n)}{|n_A(2y_n)|},-\tilde\eta\cfrac{n_B(2y_B)}{|n_B(2y_B)|} \right]$
			\\ \Xhline{3\arrayrulewidth}
			\cellcolor{Snow3} $b^\mu$ & $(0,b^-,b_\perp)$ & $(0, b_T^x,b_T^y,\tilde b^z)$
			\\ \hline
			\cellcolor{Snow3} $v^\mu$ & $(-e^{2y_B},1,0_\perp)$ & $(0,0,0,-1)$ 
			\\ \hline
			\cellcolor{Snow3} $\delta^\mu$ & $(0,b^-,0_\perp)$ & $(0,0,0,\tilde b^z)$
			\\ \hline
			\cellcolor{Snow3} $P^\mu$ & $\dfrac{m_h}{\sqrt2} (e^{y_P},e^{-y_P},0_\perp)$ & $m_h(\cosh y_{\tilde P}, 0,0,\sinh y_{\tilde P})$
			\\ \hline
		\end{tabular}
		\caption{Definition of the Collins and quasi-TMDs for quarks and gluons. The Large Rapidity and Collins TMDs are identical, except for the order of the limits and expression for $Z_{\rm UV}^{\kappa_i}$ in the first row.
		}\label{tbl:TMDs}
	\end{table*}
}

\paragraph{Large Rapidity (LR) scheme}
To facilitate deriving the factorization relation between the quasi- and Collins TMDs, Ref.~\cite{Ebert:2022fmh} introduced the Large Rapidity (LR) scheme,
\begin{align} \label{eq:tmdpdf_LR}
	&f^\LR_{g/h}(x, \bt, \mu, \zeta,y_P-y_B) \\
	&\quad 
  = \lim_{\substack{-y_B \gg 1}}\, \lim_{\eps\to0}\, Z_\uv^\LR(\eps, \mu, y_n-y_B)
	\frac{ B_{g/h}^{C\rho\sigma}(x, \bt, \eps, y_P - y_B)}{\sqrt{S^g \left[b_\perp,\eps,-\infty n_A(2y_n),-\infty n_B(2y_B)\right]}}
	\,,\nn
\end{align}
which differs from the Collins scheme only by its order of $\eps\to0$ and $y_B\to -\infty$ limits.

Using Lorentz invariance and the large $\tilde P^z$ expansion ($-y_B\gg 1$), one can demonstrate that when $\tilde\zeta=\zeta$ or $y_{\tilde P}=y_P-y_B$, the quasi- and LR TMDs are equivalent at leading order under a large rapidity expansion~\cite{Ebert:2022fmh}, i.e.,
\begin{align}\label{eq:quasiLR}
	\tilde f_{g/h}(x, \bt, \mu, \tilde\zeta, x \tilde P^z) &= f^\LR_{g/h}(x, \bt, \mu, \tilde \zeta, y_P - y_B)  + {\cal O}(y_B^k e^{y_B}) \,,
\end{align}
where ${\cal O}(y_B^k e^{y_B})$ are exponentially suppressed contributions.

\section{Calculation of one-loop gluon matching coefficient}
\label{sec:match}

We now compute the one-loop quasi-to-Collins matching coefficient $C_g$ for gluons.
To do so, we actually carry out a simpler equivalent matching calculation: between the LR and Collins schemes.
In the course of proving the quasi-to-Collins factorization formula in \eq{factorization_statement}, \refcite{Ebert:2022fmh} showed that the quasi-to-Collins and LR-to-Collins matching coefficients are identical. 
Note that this matching calculation focuses on beam functions: the quasi-soft function has been chosen to reproduce the Collins soft function as $|\tilde \eta|\to \infty$. 

Let us compute the matrix element of the Collins beam function for a free external gluon state with momentum $p^\mu=(p^+,0,0_\perp)$, transverse polarization vectors $\epsilon^{\alpha}_\perp $ and $\epsilon^\beta_\perp $, as well as color indices $a$ and $ b$. 
We work in Feynman gauge and dimensional regularization with space-time dimension $d=4-2\eps$.
As we are interested in the leading power gluon TMDs, the Lorentz indices $\rho$, $\sigma$ are both transverse,
and for both the LR and Collins schemes $\mu=\nu=+$.
In our calculation we leave the choice of $\rho$ and $\sigma$ unspecified, to make clear that our result applies to all spin-dependent gluon TMDs.

\begin{figure}[t]
	\centering
	\begin{subfigure}{0.45\textwidth}
		\raisebox{-13.5ex}{\includegraphics[width=\textwidth]{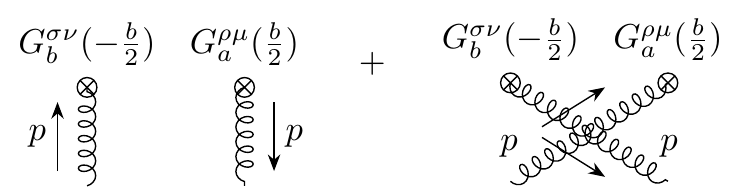}}
		\caption{}
		\label{fig:gga}
	\end{subfigure}
	\begin{subfigure}{0.4\textwidth}
		\includegraphics[width=\textwidth]{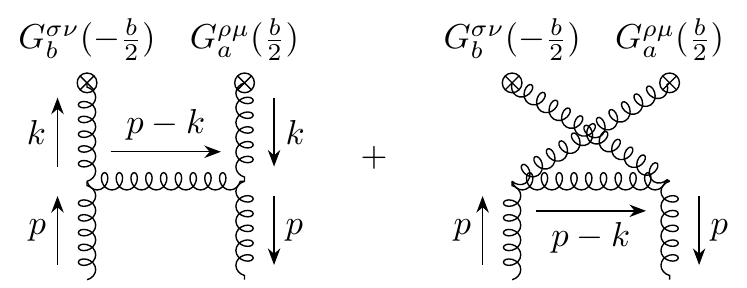}
		\caption{}
		\label{fig:ggb}
	\end{subfigure}

	\begin{subfigure}{0.8\textwidth}
		\includegraphics[width=\textwidth]{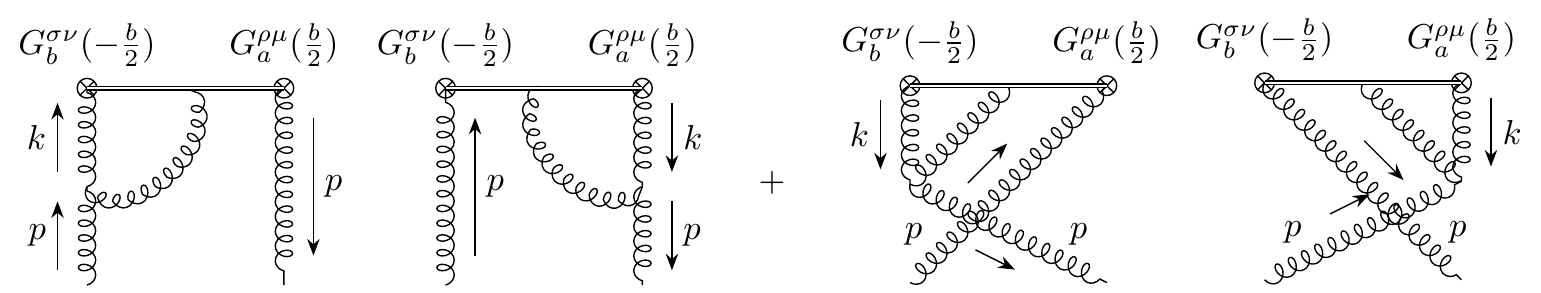}
		\caption{}
		\label{fig:ggc}
	\end{subfigure}
	
	\begin{subfigure}{0.8\textwidth}
		\includegraphics[width=\textwidth]{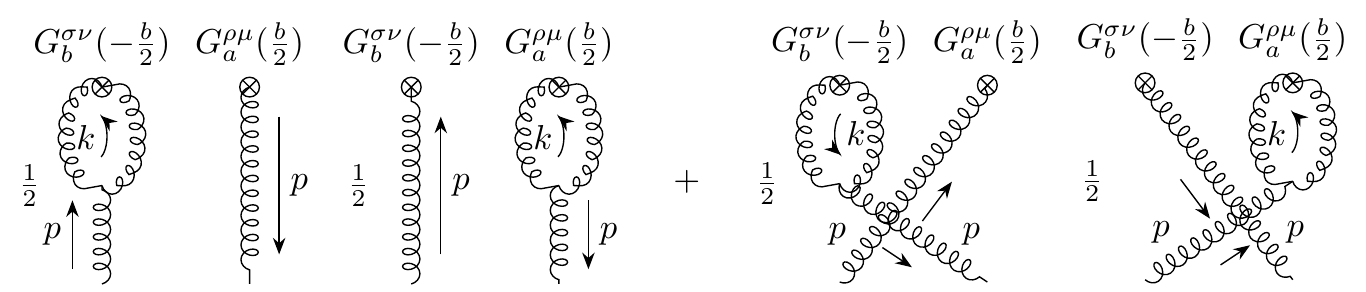}
		\caption{}
		\label{fig:ggd}
	\end{subfigure}
	
	\begin{subfigure}{\textwidth}
		\includegraphics[width=\textwidth]{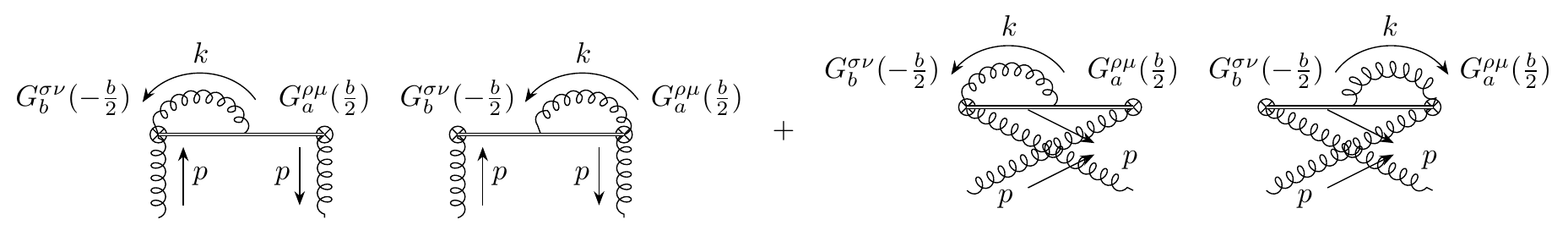}
		\caption{}
		\label{fig:gge}
	\end{subfigure}

	\begin{subfigure}{0.8\textwidth}
		\includegraphics[width=\textwidth]{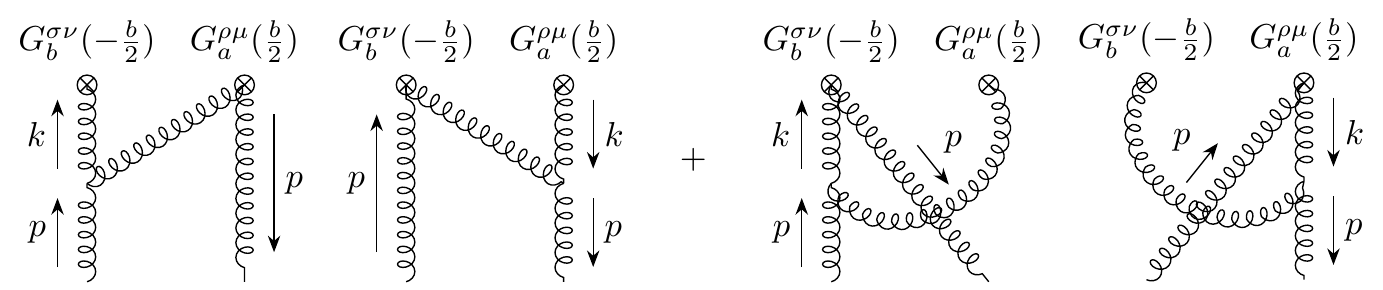}
		\caption{}
		\label{fig:ggf}
	\end{subfigure}

	\begin{subfigure}{0.45\textwidth}
		\includegraphics[width=\textwidth]{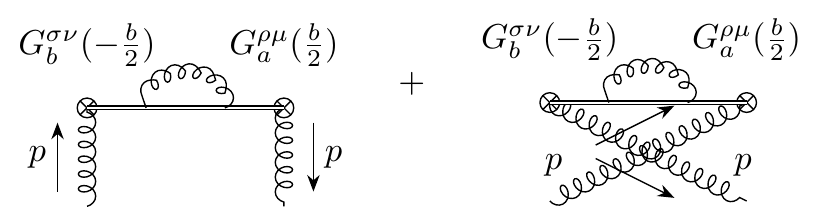}
		\caption{}
		\label{fig:ggg}
	\end{subfigure}
	\begin{subfigure}{0.45\textwidth}
		\includegraphics[width=\textwidth]{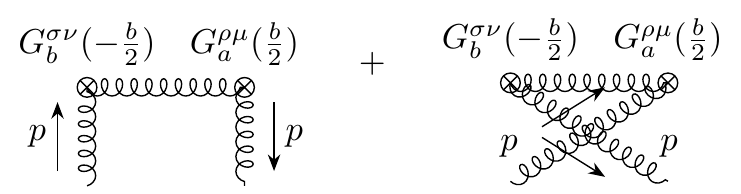}
		\caption{}
		\label{fig:ggh}
	\end{subfigure}
	\caption{The tree-level (a) and one-loop level (b--h) diagrams contributing to gluon beam function in a free gluon state. Since the external gluons are transversely polarized, they cannot be attached to the Wilson lines in the $n_B$ direction. 
     While depicted in these diagrams as a straight line, the beam function Wilson line actually traverses an infinitely long staple-shaped path, as shown in \fig{generic_staples}a.
	}
	\label{fig:diagrams_gg}
\end{figure}

At tree-level, we must compute the diagrams in \fig{gga}, which give
\begin{align}
\Corr_{g/g}^{+\rho+\sigma(0)}(b^\mu)	&= \delta^{ab} (p^+)^2 \left[ g^{\rho \beta}g^{\sigma \alpha} e^{ip\cdot b} + (\alpha\leftrightarrow \beta, p\to -p) \right]\,,
\end{align}
where the second term in the square bracket comes from the exchange symmetry of bosonic particles. 
Taking a Fourier transform $b^z \to xp^z$, the tree-level gluon beam function becomes
\begin{align}
	x B_{g/g}^{C\rho\sigma(0)}(x, \bt) &= \delta^{ab} \left[ g^{\rho \beta}g^{\sigma \alpha} \delta(1-x) + (\alpha\leftrightarrow \beta, x\to -x) \right]\,,
\end{align}
where we moved the normalization factor $1/x$ to the left hand side for clarity.
We emphasize that to find the matching coefficient, we need only examine terms proportional to $\delta(1-x)$ since the factorization relation does not involve a convolution in $x$.

The one-loop Feynman diagrams that contribute to the beam function are shown in \fig{diagrams_gg}.
According to Ref.~\cite{Ebert:2022fmh}, only diagrams containing a rapidity divergence contribute to the matching, significantly simplifying our task. 
Only  \figs{ggc}{gge} could exhibit rapidity divergences, as the other diagrams have Lorentz-covariant loop integrals, which remain unaffected by the large rapidity limit $y_B\to-\infty$ (except for the transformation of tensor structures). 
Using the Feynman rules in \app{rules}, we  express \figs{ggc}{gge} as
\begin{align}\label{eq:diagc1}
	\Corr_{g/g, c}^{+\rho+\sigma(1)}(b^\mu) &=(n_b)_{\mu}(n_b)_{\nu}\, \iota^\epsilon \!\!\int\!\! {d^dk \over (2\pi)^d}  (-ik^\sigma g^{\nu\delta} + ik^\nu g^{\sigma \delta})\delta^{fd}e^{ik\cdot {b\over 2}}\ (ip^\rho g^{\mu\beta} - ip^\mu g^{\rho\beta})\delta^{bc}e^{ip\cdot {b\over 2}}\nn\\
	&\qquad \times (g f^{fea})\left[g^{\delta\gamma}(p-2k)^\alpha + g^{\gamma\alpha}(k-2p)^\delta + g^{\alpha\delta}(p+k)^\gamma \right] \nn\\
	&\qquad \times g f^{edc}\int ds\ \gamma'_\gamma(s) e^{-i(p-k)\cdot \gamma(s)}{i\over k^2+i0} {i\over (p-k)^2+i0} 
	\nn\\
	&\quad+ (n_b)_{\mu}(n_b)_{\nu}\, \iota^\epsilon \!\!\int\!\! {d^dk \over (2\pi)^d}  (ik^\rho g^{\mu\delta} - ik^\mu g^{\rho \delta})\delta^{fc}e^{ik\cdot {b\over 2}}\ (-ip^\sigma g^{\nu\alpha} + ip^\nu g^{\sigma\alpha})\delta^{ad}e^{ip\cdot {b\over2}}\nn\\
	&\qquad \times (g f^{bef})\left[g^{\delta\gamma}(p-2k)^\beta + g^{\gamma\beta}(k-2p)^\delta + g^{\beta\delta}(p+k)^\gamma \right] \nn\\
	&\qquad \times g f^{edc}\int ds\ \gamma'_\gamma(s) e^{i(p-k)\cdot \gamma(s)}{i\over k^2+i0} {i\over (p-k)^2+i0}
 \,,
\end{align}
and
\begin{align} \label{eq:diage1}
	\Corr_{g/g, e}^{+\rho+\sigma(1)}(b^\mu) &= (n_b)_{\mu}(n_b)_{\nu} \, \iota^\epsilon \!\! \int\!\! {d^dk \over (2\pi)^d}  (g f^{dae})(g^{\nu\alpha}g^{\sigma\gamma} - g^{\nu\gamma}g^{\sigma\alpha})e^{i(p-k)\cdot {b\over2}}\nn\\
	&\qquad \times (ip^\rho g^{\mu\beta} - ip^\mu g^{\rho\beta})\delta^{bc}e^{ip\cdot {b\over2}} g f^{ecd}\int ds\ \gamma'_\gamma(s) e^{-ik\cdot \gamma(s)}{i\over k^2+i0} \nn\\
	&\quad +(n_b)_{\mu}(n_b)_{\nu} \int {d^dk \over (2\pi)^d}  (g f^{ceb})(g^{\mu\gamma}g^{\rho\beta} - g^{\mu\beta}g^{\rho\gamma}) e^{-i(k-p)\cdot {b\over2}}\nn\\
	&\qquad \times (-ip^\sigma g^{\nu\alpha} + ip^\nu g^{\sigma\alpha})\delta^{ad}e^{ip\cdot {b\over2}} g f^{ecd}\int ds\ \gamma'_\gamma(s) e^{ik\cdot \gamma(s)}{i\over k^2+i0} \,,
\end{align}
where the Wilson line takes a staple-shaped path:
\begin{align}
\gamma(s) : \frac{b}{2} ~\to~ \frac{b}{2}  - {b^- n_b\over 2} - \infty n_B
~\to~ -\frac{b}{2}  + {b^- n_b\over 2} -\infty n_B
~\to -\frac{b}{2}\,,
\end{align}
and we have suppressed the contributions from diagrams with external gluon lines exchanged $(\alpha\leftrightarrow \beta, p\to -p)$ for conciseness. We also use a coupling in the $\MSbar$ scheme, which introduced the factors $\iota^\epsilon = (e^{\gamma_E}/(4\pi))^\epsilon$. 

Simplifying the above,
\begin{align}
	\Corr_{g/g, c}^{+\rho+\sigma(1)}
	&= -\!2i g^2 C_A p^{+2}g^{\alpha\sigma}g^{\beta\rho}
    \iota^\epsilon \!\!\!\int\!\! {d^dk \over (2\pi)^d}{k^+\! (e^{ik\cdot b} \!-\! e^{ip\cdot b})\over k^2(p-k)^2}
	\Big[{1\over n_B\!\cdot\! (p\!-\!k)\!+\! i \delta}  + {1\over n_B\!\cdot\! (p\!-\!k) \!-\! i \delta}  \Big]\nn\\
	&+ 2ig^2C_A e^{2y_B}p^+ \iota^\epsilon \!\!\int\!\! \frac{d^dk}{(2\pi)^d}{e^{ik\cdot b} - e^{iP\cdot b} \over k^2(p-k)^2}
	\left[{k^\alpha k^\sigma g^{\beta\rho}\over n_B\cdot (p-k) + i \delta}  + {k^\beta k^\rho g^{\alpha\sigma}\over n_B\cdot (p-k) - i \delta}  \right]\nn\\
	& - 2ig^2C_A\ p^+g^{\alpha\sigma}g^{\beta\rho} \iota^\epsilon \!\! \int\!\! {d^dk \over (2\pi)^d}{k^+ \over k^2(p-k)^2}\left(e^{ik\cdot b} - e^{ip\cdot b}\right)\,,
\label{eq:diagc2}\\
	\Corr_{g/g, e}^{+\rho+\sigma(1)}
	&= ig^2 C_A e^{2y_B} p^+ g^{\alpha\sigma}g^{\beta\rho} \iota^\epsilon \!\! \int\!\! {d^dk \over (2\pi)^d}  {1\over k^2} \left(e^{-ik\cdot b} \!-\! 1\right)e^{ip\cdot b}
	\left[{1\over n_B\!\cdot\! k \!+\! i \delta}  + {1\over n_B\!\cdot k \!-\! i \delta}  \right] \,,
\end{align}
where we obtain the Wilson line propagators $[n_B\cdot (p-k) \pm i\delta]^{-1}$ by integrating over the path parameter $s$.
Note that in the LR and Collins scheme that the staple Wilson line has infinite extent. 
We introduce the infinitesimal imaginary part $\pm i\delta$ to ensure that the path integral contribution from $s=-\infty$ properly vanishes in Feynman gauge. 

In the $y_B\to -\infty$ limit, the rapidity divergence is in the first line of \eq{diagc2}. 
The second line is suppressed by a factor of $e^{2y_B}k_T^2/(p^+)^2$ compared to the first line, as $k_T=|k_\perp| \ll k^+, p^+$ in the power counting, and $\alpha,\beta,\rho,\sigma = \perp$. 
The last line in \eq{diagc2} is Lorentz covariant and so is not affected by $y_B\to-\infty$; it thus does not contribute to the matching. 

After taking a Fourier transform $b^-\to xp^+$, the first line of \eq{diagc2} becomes
\begin{align}\label{eq:diagc3}
	 x &B_{g/g, c}^{C\rho\sigma(1)}(x, \bt, \eps, y_P-y_B)\\
	=&- 2i g^2 C_A\, p^+ g^{\alpha\sigma}g^{\beta\rho}  \iota^\epsilon \!\! \int\!\! {d^dk \over (2\pi)^d}{k^+ e^{-i\vec{k}_T\cdot \vec{b}_T} \over k^2(p-k)^2} \left[{1\over n_B\cdot (p-k) +i \delta}  + {1\over n_B\cdot (p-k) - i \delta}  \right] \nn\\
	&\hspace{5cm}\times \left[\delta(k^+-xp^+) - \delta(p^+-xp^+)\right] \nn\\
	& - 2i g^2 C_A\ g^{\alpha\sigma}g^{\beta\rho}\delta(1-x) \iota^\epsilon \!\!\int\!\! {d^dk \over (2\pi)^d}{k^+(e^{-i\vec{k}_T\cdot \vec{b}_T} - 1)\over k^2(p-k)^2}\left[{1\over n_B\cdot (p\!-\!k) \!+\!i \delta}  + {1\over n_B\cdot (p\!-\!k) \!-\! i \delta}  \right]
  .\nn
\end{align}
The first term of \eq{diagc3} is a plus function in $x$; it is unaffected by the $y_B\to -\infty$ limit as long as $x\neq1$. The second term contains the contribution to the matching coefficient. 
Likewise, after a Fourier transform $b^-\to xp^+$,  the second line of \eq{diagc2}  becomes
\begin{align}\label{eq:diage2.1}
	 x B&_{g/g, e}^{C\rho\sigma(1)}(x, \bt, \eps, y_P-y_B)\\
	=& ig^2 C_A\ e^{2y_B} g^{\alpha\sigma}g^{\beta\rho} \iota^\epsilon \!\! \int\!\! {d^dk \over (2\pi)^d}  {1\over k^2}\left[{1\over n_B\cdot k +i \delta}  + {1\over n_B\cdot k - i \delta}  \right]\left(e^{i\vec{k}_T\cdot \vec{b}_T} - 1\right)\nn\\
	&\hspace{4.5cm}\times  \left[ \delta(k^+-(1-x)p^+) - \delta(p^+-xp^+)\right]\nn\\
	& + ig^2 C_A\ e^{2y_B} g^{\alpha\sigma}g^{\beta\rho}{\delta(1-x)\over p^+} \iota^\epsilon \!\! \int\!\! {d^dk \over (2\pi)^d}  {1\over k^2}\left[{1\over n_B\cdot k +i \delta}  + {1\over n_B\cdot k - i \delta}  \right]\left(e^{i\vec{k}_T\cdot \vec{b}_T} - 1\right) \,.\nn
\end{align}
The first term is again a plus function. The second term vanishes under parity transforms $k^\pm\to -k^\pm$. Therefore, the diagrams in \fig{gge} do not contribute to the matching.

All that remains to compute is the term proportional to $\delta(1-x)$ in \eq{diagc3}. After a change of variables $k\to p-k$, the next step is to evaluate the integral
\begin{align}
	I &=  \iota^\epsilon \!\! \int\!\! {d^dk \over (2\pi)^d}{p^+-k^+\over k^2(p-k)^2}\left(e^{i\vec{k}_T\cdot \vec{b}_T} - 1\right) \left[{1\over n_B\cdot k +i \delta}  + {1\over n_B\cdot k - i \delta}  \right] 
\,.
\end{align}
To simplify the integral we can make a change of variable $k^-\to e^{-y_B}k^-$ and $k^+ \to e^{y_B}k^+$ which corresponds to a boost to equal-time with $k^z=(k^+-k^-)/\sqrt2$, and yields
\begin{align} \label{eq:finalI}
	I &= \frac{ \iota^\epsilon}{\sqrt{2}} \int {d^{d}k\over (2\pi)^d} {p'^+-k^+\over (k^2 +i0 ) \left[(p'-k)^2 +i0\right]} \left(e^{i\vec{k}_T\cdot \vec{b}_T} -1\right)   \left[{1\over k^z -i \delta}  + {1\over k^z + i \delta}  \right]\,,
\end{align}
where $p'^\mu = (p^+e^{-y_B},0,0_\perp)$.
Note that the result involves a principal value integral for $k^z$,
\begin{align}
 {1\over 2} \left({1\over k^z +i\delta} + {1\over k^z - i\delta}\right)
 = {\rm PV}\left( {1\over k^z}\right) \,.
\end{align}
We must evaluate the integral in \eq{finalI} with two different orders of limits.

\paragraph{Taking $y_B\to -\infty$ before $\eps\to0$.} We first perform a contour integration in the complex $k^0$ plane, then we integrate over $k^z$ to obtain:
\begin{align}
	I&={i\over4}  \iota^\epsilon \!\!\int\!\! {d^{d-2}k_T\over (2\pi)^{d-1}}  \left(e^{i\vec{k}_T\cdot \vec{b}_T} -1\right) \left[ \left({2\over k_T^2} + {1\over p'^2_z}\right) 
	\frac{	\ln \left(  \frac{k_T^2 + 2p'_z\sqrt{k_T^2+p'^2_z} + 2p'^2_z}{k_T^2}   \right)}
	{\sqrt{1+k_T^2/p'^2_z}}
	- {4\over k_T^2} \right]\,.
\end{align}
Here $p^{\prime\,2}_z = p^{+2} e^{-2y_B}/2$.
The integrand diverges in the $p'^z\to\infty$ or $y_B\to-\infty$ limit, so by taking the limit first we carrying out the power expansion in $p'^z$ before integrating over $k_\perp$, and keeping the $\ln(p'^z)$ singularity, which leads to
\begin{align} \label{eq:I1}
		I_{y_B\to-\infty}&={i\over 4\pi} \iota^\epsilon \!\!\int\!\! {d^{d-2}k_T\over (2\pi)^{d-2}} {1\over k_T^2} \left[-2 + \ln{4p'^2_z \over k_T^2} \right]\left(e^{i\vec{k}_T\cdot \vec{b}_T} -1\right)\,,\\
		&= {i\over (4\pi)^2}\left[{1\over \eps^2} + {1\over \eps} \Big(2+\ln {\mu^2\over 4p'^2_z}\Big) - {1\over2}\ln^2\Big({b_{T}^2\mu^2\over b_0^2}\Big)  + \ln\Big({b_{T}^2\mu^2\over b_0^2}\Big)\Big(2+\ln {\mu^2\over 4p'^2_z}\Big) - {\pi^2\over 12}\right]
   \,,\nn
\end{align}
where $b_0=2e^{-\gamma_E}$, and the $1/\eps^2$ and $1/\eps$ poles are UV divergences. In the $\MSbar$ scheme we carry out the renormalization by simply subtracting all the $1/\eps$ poles.

\paragraph{Taking $\eps\to0$ before $y_B\to -\infty$.} Here, we can directly integrate:
\begin{align}\label{eq:I2}
	I_{\eps\to0}
	&={\iota^\epsilon \over2} \int\!\! {d^{d}k\over (2\pi)^d} {2p'^z-k^0\over (k^2 +i0 ) \left[(p'-k)^2 +i0\right]} \left(e^{i\vec{k}_T\cdot \vec{b}_T} -1\right)   \left[{1\over k^z -i \delta}  + {1\over k^z + i \delta}  \right]\nn\\
	&\qquad - \iota^\epsilon\!\! \int\!\! {d^{d}k\over (2\pi)^d} {1\over (k^2 +i0 ) \left[(p'-k)^2 +i0\right]} \left(e^{i\vec{k}_T\cdot \vec{b}_T} -1\right) \nn\\ 
	&= {i\over (4\pi^2)} \left[{1\over \eps} -{1\over2}\ln^2{(4p'^z b_T)^2 \over b_0^2} + \ln{(4p'^z b_T)^2 \over b_0} + \ln({b_{T}^2\mu^2\over b_0^2})- 2\right]
 \,.
\end{align}
Again the $1/\eps$ pole is a UV divergence, and is subtracted for the $\MSbar$ renormalized result.

Finally we note that the one-loop Collins soft function is the same for quarks and gluons, so can be obtained from the quark result~\cite{Ebert:2019okf}  with the replacement of $C_F\to C_A$,
\begin{align}
	S_{C}^{g,(1)}(b_T,\eps,2y_n,2y_B) 
&= {\alpha_sC_A\over 2\pi} \iota^\epsilon {\Gamma(-\eps) \over 4^{\eps}} (b_T^2\mu^2)^{\eps}\left[2(y_n\!-\!y_B) {1\!+\!e^{2(y_B-y_n)} \over 1\!-\! e^{2(y_B-y_n)} } \!-\! 2\right]  \,.
\end{align}
Here we use the shorthand $S_C^g(b_T,\eps, y_A,y_B) 
 = S^g[b_\perp, \eps, -\infty \nA(y_A), -\infty \nB(y_B)]$. This result is not affected by the order of $\eps\to0$ and $y_B\to -\infty $ limits, i.e.,
\begin{align}
	\lim_{\substack{\eps\to0 \\ -y_B\gg 1}}S_{C}^{g,(1)}(b_T,\eps,2y_n,2y_B) 
 &= {\alpha_sC_A\over 2\pi}\left({1\over \eps} + \ln{b_T^2\mu^2\over b_0^2}\right)[2-2(y_n-y_B)] \,.
\end{align}
Therefore this result cancels out for the matching between the LR and Collins schemes.

\paragraph{Matching coefficient.}  The difference between the renormalized LR and Collins scheme results in \eqs{I1}{I2} gives
\begin{align}\label{eq:diff}
	I_{\eps\to0}^{\rm ren} \!-\! I_{y_B\to-\infty}^{\rm ren}
  &=-{i\over (4\pi^2)} \left[
 {1\over2}\ln^2 {\mu^2\over 4p'^2_z} + \ln {\mu^2\over 4p'^2_z} + 2 - {\pi^2\over 12}\right]\,,
\end{align}
which is purely of ultraviolet origin, and hence is perturbative at a scale $\mu^2\sim 4 p_z'^2$.
Thus the one-loop matching coefficient between the LR and Collins TMD  is given by
\begin{align}
	C_g(p'^z,\mu) &= 1-{\alpha_s(\mu) C_A\over 4\pi}\biggl[ \ln^2 {\mu^2\over 4p'^2_z} + 2\ln {\mu^2\over 4p'^2_z} + 4 - {\pi^2\over 6}\biggr]\,.
\end{align}
Note that here $4p'^2_z=2(p^+)^2e^{-2y_B}$. For a hadron with momentum $P$, we must replace $p^+\to xP^+$, so that
\begin{align}
	4p'^2_z \longrightarrow 2(xP^+)^2e^{-2y_B} \approx 4x^2m_h^2 \sinh^2(y_P-y_B)  = \zeta_\LR
\end{align}
in the limit of $-y_B\gg 1$. Therefore, identifying $\zeta_\LR= (2x \tilde P^z)^2 $, we find that the final gluon matching coefficient for the quasi-TMD is:
\begin{align}
	C_g(x\tilde P^z,\mu) &= 1-{\alpha_sC_A\over 4\pi}\biggl[ \ln^2 {\mu^2\over (2x\tilde P^z)^2} + 2\ln {\mu^2\over (2x\tilde P^z)^2} + 4 - {\pi^2\over 6}\biggr]
 \,.
\end{align}
Interestingly this result is identical to the quark result $C_q$~\cite{Ebert:2019okf}, up to changing Casimirs from $C_F\to C_A$. This occurs because the entire contribution to the matching comes from the part of the Wilson line diagrams associated with the rapidity divergence.
The well-known (generalized) Casimir scaling of the cusp anomalous dimension and CS kernel have been validated to four-loop order~\cite{Korchemsky:1987wg,Moch:2004pa,Henn:2019swt,vonManteuffel:2020vjv,Moult:2022xzt,Duhr:2022yyp}. 
The fact that the full one-loop matching and cusp-anomalous dimension obey Casimir scaling, implies that the full tower of next-to-leading-logarithms for $C_g$ will obey Casimir scaling as well~\cite{Ebert:2022fmh}. 
This hints at a relationship for quasi-TMD matching coefficients at higher orders.

\section{Conclusion}

In this paper, we calculated the one-loop gluon matching coefficient $C_g$ in the factorization formula relating quasi and Collins TMDs, marking an important step towards calculating gluon TMDs from lattice QCD. As the matching coefficient is independent of spin structure and there is no quark-gluon mixing, our result enables lattice calculation of all eight leading-twist gluon TMDs. Given the encouraging first results for quark TMDs, it is reasonable to expect that lattice QCD will provide highly useful nonperturbative input for gluon TMDs and their Collins-Soper evolution, neither of which has yet been extracted from experiments.

\section*{Acknowledgments}
This work is dedicated to Markus Ebert, who has become more industrious. 
This work was supported by the U.S. Department of Energy, Office of Science, Office of Nuclear Physics, from DE-SC0011090, DE-AC02-06CH11357 and within the framework of the TMD Topical Collaboration. I.S. was also supported in part by the Simons Foundation through the Investigator grant 327942.  S.T.S. was partially supported by the U.S. National Science Foundation through a Graduate Research Fellowship under Grant No. 1745302. Y.Z. is partially supported by an LDRD initiative at Argonne National Laboratory under Project~No.~2020-0020.

\appendix

\section{Lightcone coordinate conventions}
\label{app:notation}

We work in a frame where the hadron momentum $P$ is close to the lightlike unit vectors
\begin{align} \label{eq:na_nb}
	\na^\mu = \frac{1}{\sqrt2} (1, 0, 0, 1) \,,\qquad \nb^\mu = \frac{1}{\sqrt2} (1, 0, 0, -1)
	\,,\end{align}
which obey $\na^2 = \nb^2 = 0$ and $\na \cdot \nb = 1$.
We can decompose a four-vector $p^\mu$ is 
\begin{align}
	p^\mu = (p^+, p^-, p_\perp) = p^+ \na^\mu + p^- \nb^\mu + p_\perp^\mu
	\,,\end{align}
where $p^\pm = (p^0 \pm p^z) / \sqrt2$
and $p_\perp^\mu = (0, p^x, p^y, 0) = (0, \pt, 0)$.
Here $p_\perp^\mu$ is a Minkowski vector, and $\pt$ is the corresponding Euclidean vector with magnitude $p_T \equiv (\pt^{\,2})^{1/2} =  (-p_\perp^2)^{1/2}$.

\section{Feynman rules}
\label{app:rules}

The gluon field strength tensor is defined as $ig G^{\mu\nu}_a\tau^a = \left[ D^\mu, D^\nu\right]$,
where $iD^\mu=i\partial^\mu + g A^\mu_a\tau^a$ and $\tau^a$ is the SU(3) generator in the fundamental representation.
The Feynman rules for the local, three-gluon, and gluon-Wilson line vertices are:
\allowdisplaybreaks
\begin{align}
	\raisebox{-5ex}{\includegraphics[width=4cm]{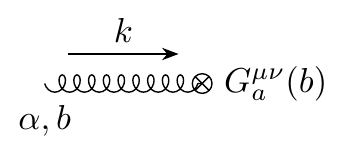}} &= (-ik^\mu g^{\nu \alpha} + ik^\nu g^{\mu\alpha})\delta^{ab}e^{-ik\cdot b}\,,\nn\\
	\raisebox{-5ex}{\includegraphics[width=4.5 cm]{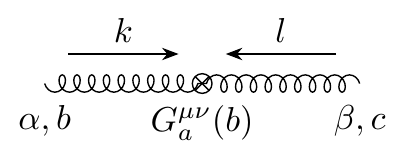}}& = g f^{abc} \left[g^{\mu\alpha}g^{\nu\beta} - g^{\mu\beta}g^{\nu\alpha}\right]e^{-i(k+l)\cdot b}\,,
\nn\\
	\raisebox{-10ex}{
    \includegraphics[width=2.8cm]{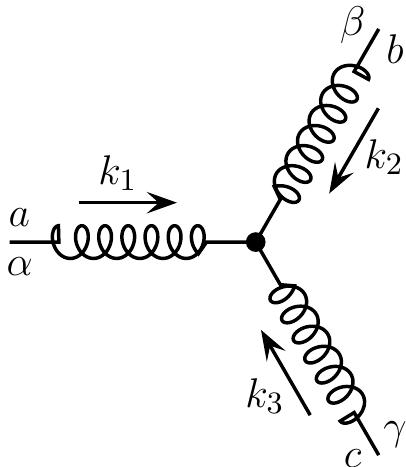}}\qquad \quad
	&= gf^{abc}\Big[g^{\alpha\beta}(k_1-k_2)^\gamma  + g^{\beta\gamma}(k_2-k_3)^\alpha + g^{\gamma\alpha}(k_3-k_1)^\alpha\Big]
	\,,\nn\\
	\raisebox{-7ex}{
    \includegraphics[width=3.0cm]{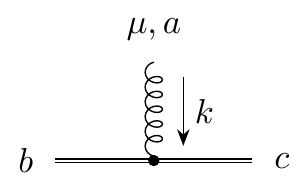}}\qquad
	&= gf^{abc} \gamma'(s)_\mu e^{-i k\cdot \gamma(s)}
\,.
\end{align}

\bibliographystyle{JHEP}
\bibliography{gluon}

\end{document}